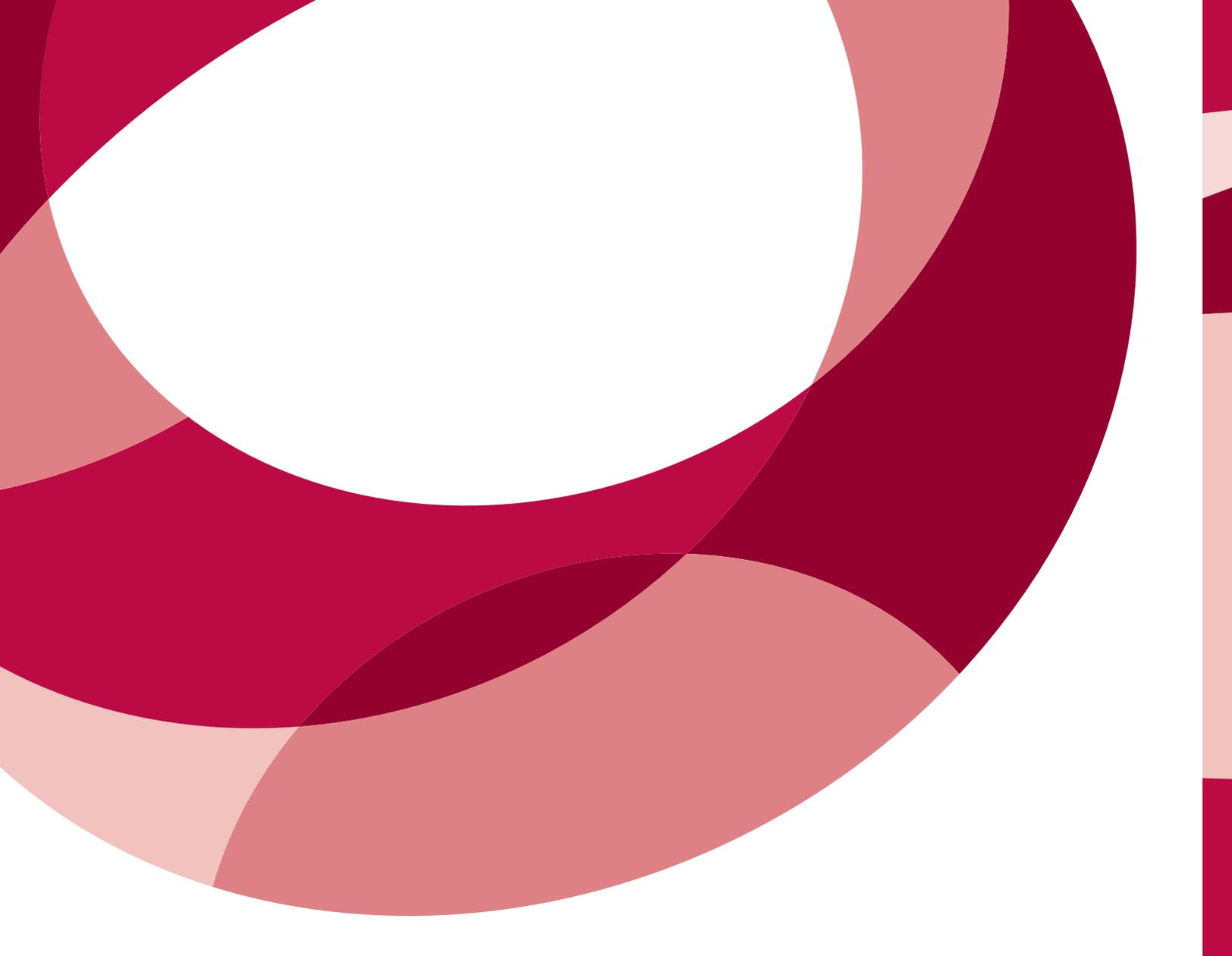

# 5 Year Update to the Next Steps in Quantum Computing Report

The material is based upon work supported by the National Science Foundation under Grant No. 1734706. Any opinions, findings, and conclusions or recommendations expressed in this material are those of the authors and do not necessarily reflect the views of the National Science Foundation.

# 5 Year Update to the Next Steps in Quantum Computing Report

## Workshop
## May 18 - 19, 2023

**Organizers**

Kenneth Brown, Duke University
Fred Chong, University of Chicago
Kaitlin N. Smith, Northwestern University and Infleqtion
Thomas M. Conte, Georgia Institute of Technology and Community Computing Consortium

**With Support From:**

Catherine Gill, Computing Community Consortium
Maddy Hunter, Computing Community Consortium
Taniya Ross-Dunmore, Computing Research Association
Ann Schwartz, Computing Community Consortium

**Authors:**
Kenneth Brown, Duke University
Fred Chong, University of Chicago
Kaitlin N. Smith, Northwestern University and Infleqtion
Thomas M. Conte, Georgia Institute of Technology and Community Computing Consortium
Austin Adams, Georgia Institute of Technology
Aniket Dalvi, Duke University
Christopher Kang, University of Chicago
Josh Viszlai, University of Chicago

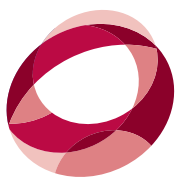
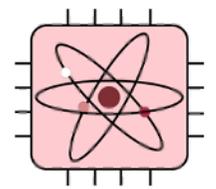

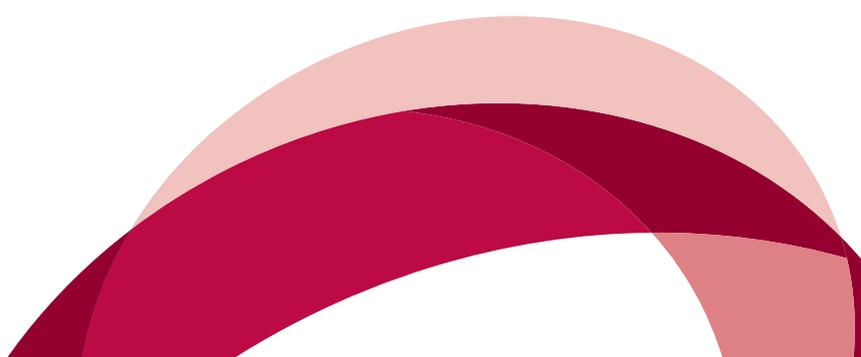

## Introduction

It has been 5 years since the Computing Community Consortium (CCC) Workshop on Next Steps in Quantum Computing, and significant progress has been made in closing the gap between useful quantum algorithms and quantum hardware. Yet much remains to be done, in particular in terms of mitigating errors and moving towards error-corrected machines. As we begin to transition from the Noisy-Intermediate Scale Quantum (NISQ) era to a future of fault-tolerant machines, now is an opportune time to reflect on how to apply what we have learned thus far and what research needs to be done to realize computational advantage with quantum machines.

Quantum computation promises to revolutionize the way we perform scientific and mathematical calculations by using quantum correlations to create algorithmic shortcuts that are inaccessible to standard computers. These algorithmic shortcuts break standard security protocols for the internet (Shor, 2006, pp. 303–332) and have forced cryptographers to develop quantum resistant methods. Quantum computers (QCs) should be ideal for simulating quantum chemistry and physics problems. Although quantum chemistry may sound esoteric, calculations on current supercomputers are used to design molecules for pharmaceuticals and to understand catalysts, with applications including battery design, energy efficient fertilizer production, and carbon sequestration. A large fraction of supercomputer time at national laboratories is currently used for quantum mechanics problems, such as those in nuclear physics or material science, and QCs promise to dramatically increase the scope and lower the cost of these simulations. More speculatively, if scalable QCs can improve optimization and machine learning, this would have major economic and technological implications.

In the last 5 years, the United States with the National Quantum Initiative has made quantum technologies and quantum information science (QIS) a priority, and Quantum Leap is one of the NSF's 10 Big Ideas. In the United States, there is a nascent QC industry with active participation from technology giants, such as Google, Microsoft, IBM, and Amazon as well as a vibrant start-up community. Despite this effort and funding, the reason that QCs do not yet provide large scientific and societal benefits is that quantum components are noisy and have a limited number of qubits and gates (Fig. 1).

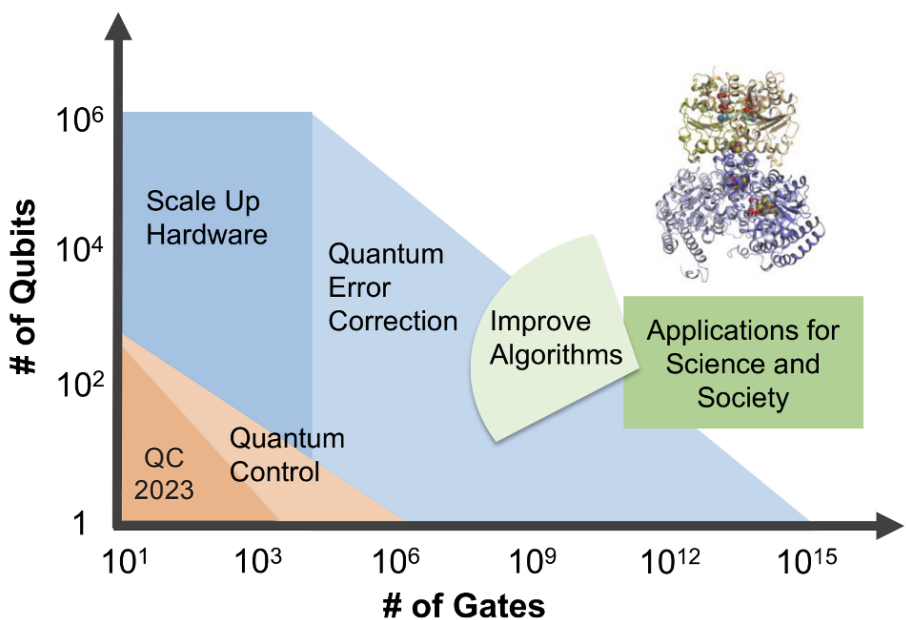

**Fig 1.** Quantum computers in 2023 have a limited number of qubits and are able to perform a limited number of gates. In order to use quantum computers to study challenging scientific problems (e.g., the quantum dynamics of metal-organic centers of enzymes), we need to be able to perform many more operations.
*(Image by K. Brown and E. Edwards)*
11

This report examines 5 critical areas of future research that will build upon the last 5 years of progress and help us move beyond the current state of noisy quantum machines. These areas are:

### 1) Technologies and Architectures with a View Towards Scaling

Scalable architectures demand that larger systems yield lower computational error and decreased per qubit costs, and reaching practical quantum computation will require creativity and cross-discipline collaboration within academia and industry to produce technological innovation that permeates the quantum compute stack. Further, improved models that are faithful to the dynamics of actual systems will help push progress forward by defining the practical constraints we must consider when making theoretical quantum systems a reality.

### 2) Applications and Algorithms

There is a clear need for more applications and algorithms with practical quantum advantage. This requires both producing near-term applications with demonstrated experimental advantage and continuing to develop keystone applications which have strong theoretical evidence of advantage. To facilitate these goals, we recommend reducing resource requirements of keystone applications, exploring near-term applications via domain integration, and benchmarking hardware to enable algorithm development.

### 3) Fault Tolerance and Error Mitigation

QCs are limited by noise. In the near-term, error mitigation will reduce application noise and quantum error correction (QEC) demonstrations will inform future QC design. Large scale quantum computation will require error correction and fault tolerance. Current developments in QEC codes present opportunities for co-design of quantum architectures. Systems that combine fault-tolerant principles and error-mitigation methods can serve as a bridge between current systems and future large-scale QCs.

### 4) Hybrid Quantum-Classical Systems: Architectures, Resource Management, and Security

Quantum hardware will likely be advantageous on specialized computations, and the solution of most practical problems will require a hybrid solution with substantial classical computation in cooperation with a quantum kernel. The organization of these hybrid systems and the hybrid algorithms that run on them will be key areas of research. Classical computation for quantum circuit optimization, simulation, and verification will also be key enablers. Finally, an emerging concern is the secure design of quantum systems in the face of potential vulnerabilities.

### 5) Tools and Programming Languages

The tools for quantum programming are still relatively new. Quantum programming today requires a deep knowledge of unitary mathematics and its associated linear algebra. Even with this knowledge, well-known algorithms are non-intuitive to newcomers, and, new algorithms are difficult to reason about even for quantum experts. To welcome newcomers to the field, to facilitate research, and to permit scaling up to programs with quantum advantage, efficient high-level quantum programming abstractions are needed. To realize such abstractions, software engineering infrastructure is needed for compilation, verification, and simulation, both for near-term and long-term hardware.

# I. Technologies and Architectures with a View toward Scaling

## Quantum Information Technologies

| | | Qubit Coherence Time (sec) | Two-qubit Gate Fidelity | Qubits Connected | Companies | Pros | Cons |
|---|---|---|---|---|---|---|---|
| **Natural Qubits** | | | | | | | |
| 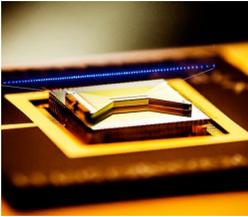 | **Trapped Ions** Electrically charged atoms, or ions, are held in place with electric fields. Qubits are stored in electronic states. Ions are pushed with laser beams to allow the qubits to interact. | >1000 | 99.9% | High | IonQ, Quantinuum, AQT Oxford Ionics, Universal Quantum | Very stable. Highest achieved gate fidelities. | Slow operation. Many lasers are needed. |
| 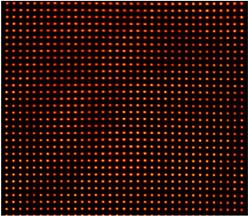 | **Neutral Atoms** Neutral atoms, like ions, store qubits within electronic states. Laser activates the electrons to create interaction between qubits. | 1 | 99.5% | Very high; low individual control | Infleqtion, Atom Computing, QuEra, Pasqal, Planqc, $M^2$ | Many qubits, 2D and maybe 3D. | Hard to program and control individual qubits; prone to noise. |
| 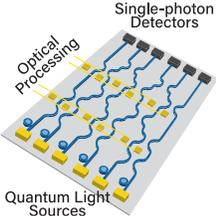 | **Photonics** Photonic qubits are sent through a maze of optical channels on a chip to interact. At the end of the maze, the distribution of photons is measured as output. | – | – | – | PsiQuantum, Xanadu | Linear optical gates, integrated on-chip. | Each program requires its own chip with unique optical channels. No memory. |
| 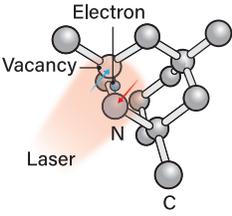 | **Diamond Vacancies** A nitrogen atom and a vacancy add an electron to a diamond lattice. Its quantum spin state, along with those of nearby carbon nuclei, can be controlled with light. | 10 | 99.2% | Low | Quantum Diamond Technologies, Quantum Brilliance | Can operate at room temperature. | Difficult to create high numbers of qubits, limiting compute capacity. |
| **Synthetic Qubits** | | | | | | | |
| 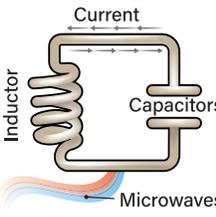 | **Superconducting Circuits** A resistance-free current oscillates back and forth around a circuit loop. An injected microwave signal excites the current into super-position states. | 0.00005 | 99.9% | High | Google, IBM, QCI, Rigett, Oxford Quantum Circuits | Can lay out physical circuits on chip. | Must be cooled to near absolute zero. High variability in fabrication. Lots of noise. |
| 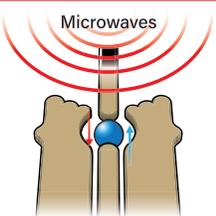 | **Silicon Quantum Dots** These "artificial atoms" are made by adding an electron to a small piece of pure silicon. Microwaves control the electron's quantum state. | 0.03 | ~99% | Very Low | HRL, Intel, SQC, Oxford Quantum Ocean, DIRAQ, Quantum Motion, EeroQ | Borrows from existing semiconductor industry. | Only a few connected. Must be cooled to near absolute zero. High variability in fabrication. |
| 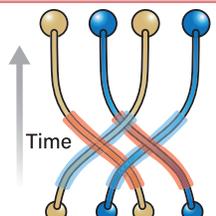 | **Topological Qubits** Quasiparticles can be seen in the behavior of electrons channeled though semiconductor structures. Their braided paths can encode quantum information. | – | – | – | Microsoft | Designed to be more robust to environmental noise. | Existence not yet confirmed. |

A summary of some of the leading quantum information technologies and their characteristics. Table modified from Gabriel Popkin, Quest for qubits. Science 354, 1090-1093(2016). DOI:10.1126/science.354.6316.10 90.

The last five years have seen an explosion of creative ideas, new experiments, and demonstrations of quantum information processing with an eye toward scaling. The key scaling goal is to architect and implement a system for which computational error decreases as the system size increases. As we look towards technological deployment, it will also be critical to devise architectures for which the cost per qubit decreases with increasing system size. The community does not currently have a clear "winner" for a single platform that scales optimally, but many promising hardware implementations exist, with efforts to improve resilience during computation in progress.

## 1.1 Quantum breakthroughs with novel approaches

Significant advances in increasing QC size and performance are desperately needed, and investment in new and risky alternative architectures has potential for high rewards. Immense activity in novel architectures has opened doors to new QEC schemes and demonstrations of some of the first logical qubits and beyond-break-even error correction (Acharya et al., 2022). Quantum systems have made great strides by exploring novel qubits, materials, and processing techniques (Place et al., 2021). In addition, fundamentally new architectures that bring us closer to utility scale are often enabled when elements and devices with individual strengths are combined heterogeneously (Stein et al., 2023). Thus, creativity in systems architecture should be fueled in the coming years to open up new opportunities in scalable technology. Groundbreaking innovation results from encouraging the community to develop novel qubits and architectures, even if they look very different from existing approaches. For example, mixing digital and continuous variable quantum computing, developing and mixing new error correction and mitigation schemes, or developing computing architectures specialized to particular tasks could substantially lower barriers to scaling. However, high-risk and potentially high-reward investments must not be made uniformly. To help understand impact potential, we need continued improvements on our tools for resource estimates. These resource estimates improve our ability to forecast where fundamentally new alternatives can have the highest return. We caution that while we can learn much from the history of classical computing, the trajectory for scaling quantum systems will likely look very different. We should not expect advances like technology convergence or standardization to happen the same manner as with classical system scaling. Flexibility is still needed as we explore potential breakthroughs in QC architectures.

## 1.2 Defining goals with realistic operation in mind

Developing technology with impact requires a clearly defined target specification during the hardware design period. As a result, scaling quantum computing technology will require developing and parametrizing models of devices, and in turn developing useful abstractions to allow for engineering, architecture, and optimization up the stack. At a basic level, those who do physics-based modeling of devices and materials should define and develop useful parameters that have predictive value in device performance, and can be used for designing new devices and architectures. One particularly interesting frontier is to address and engineer noise and loss, both to improve devices and components in a manner that is compatible with and enables scaling to large system sizes, and also to find new ways to parametrize such noise to open new opportunities in error correction, software, and algorithms to exploit features of the noise. Recent progress in developing new QEC schemes based on biased noise and erasure conversion serve as inspiration for new research directions in this space. Another critical need will be scalable methods for simulation and design, ranging from specific device-level models (such as physics-based electromagnetics and materials calculations) to more general behavioral models, similar to the approach for nested, multi-level simulation used to design classical processors.

Full-system performance simulators (where the "system" consists of the compiler, runtime, control electronics in simulation and qubit simulator) are also helpful for enabling the development of the hardware and software in tandem. Hardware/software co-design ensures the interoperability of all parts of the computational stack, enables scalability, and is another best practice leveraged from classical computing system design. System performance simulators could be useful for comparing the implementation of benchmarks on different quantum machines. These full computational stacks in simulation could also be used to test out new hardware designs (for instance in the control electronics) in simulation before building them in hardware, and could lead to the development of new quantum applications and challenge problems.

Increasing the number of qubits in a quantum system is unrealistic without the careful design of appropriate support infrastructure that includes hardware and software components. For instance, we must consider as quantum systems scale the need to place and route additional hardware that provides each qubit with adequate control. In addition, this control hardware must be accompanied by software that optimizes the conversion of quantum instructions to control signals, signal scheduling, and measured response processing. Additional hardware presents challenges when the quantum device packaging and cooling system is spatially constrained and must be done in a manner where additional noise is not injected into the system. Further, as the amount of qubit control and readout elements increase, the I/O in the interface between the QC and the classical controller could become bottlenecked, inhibiting on-the-fly quantum-classical processing. Pursuing modular architecture alleviates some of these challenges, but then it becomes essential to employ specialized mapping and compilation techniques and/or circuit knitting to enable distributed computing.

## 1.3 Bridging the gap between research communities

The interdisciplinary nature of quantum computing research requires frequent and effective communication and dissemination of research ideas across communities. An encouraging sign in the past five years is the growth of quantum presence in conferences in the area of computer architecture and design automation. Many of the project findings disseminated at these conferences are results of computer scientists and engineers working closely with experimental physicists.

Scaling up quantum computing systems necessitates broad involvement from the computer science community. For example, collaboration with the High-Performance Computing (HPC) community, which possesses extensive expertise in architecting, benchmarking, and executing large-scale computing systems, will be essential in addressing the challenges associated with large-scale quantum systems, such as optimizing hardware design, improving error correction and mitigation techniques, and enhancing overall system performance. Furthermore, as qubit processing and communication technologies continue to advance, a scalable system will involve networking multiple quantum chiplets or processors together. By engaging with the classical distributed systems and quantum network communities, we can address the challenges associated with building a robust and efficient infrastructure for large-scale quantum computing.

## 2. Applications and Algorithms

Despite the pace of development of quantum hardware, the field has not yet demonstrated an experimental advantage of quantum algorithms. While some keystone applications have strong theoretical evidence of advantage, like factoring and digital Hamiltonian simulation, these applications require fault-tolerant hardware resources that may be decades away. If realizable, these applications could solve computational problems which would otherwise be intractable, like breaking RSA or simulating molecules to chemical accuracy.

To continue growing the field, the community must accelerate the development of applications with advantage. This requires both (1) producing near-term applications with demonstrated experimental advantage and (2) continuing to develop keystone applications which have strong theoretical evidence of advantage. To facilitate these goals, we recommend:

◗ **Reducing resource requirements of keystone applications:** We hope to reduce the number of logical qubits and depth of circuits required to run keystone applications. Our priority should be working toward the first demonstration of exponential quantum advantage on a fault-tolerant device, which can be achieved via co-design of architectures, algorithms, and applications.

- Prioritize algorithms and applications requiring modest Fault-Tolerant Quantum Computing (FTQC) resources, instead of asymptotically superior algorithms with substantial resource requirements

- Translate research in near-term applications to reduce FTQC resource requirements

- **Exploring near-term applications via domain integration:**
  - Engage domain scientists to encourage interdisciplinary collaboration and explore the possibilities of near-term quantum algorithms
  - Recognize quantum-inspired algorithms as valid deliverables from QC research
- **Benchmarking hardware to enable algorithm development:** Rigorous full-stack benchmarks will facilitate algorithm debugging and execution
  - Explore standardized benchmarking frameworks to holistically evaluate the quantum stack, including domain problems, quantum algorithms, quantum platforms, and the software stack.

## 2.1 Keystone applications remain promising but decades away

Two keystone applications, factoring and Hamiltonian simulation, remain the most promising applications for QCs. In both settings, the algorithms which enable these applications are well-studied and have theoretical evidence of exponential advantage.

We recommend prioritizing research that would enable minimum viable product examples of advantage sooner. This includes architecture co-design and the design of efficient algorithms.

**Keystone applications of factoring and Hamiltonian simulation remain compelling**

Factoring served as the impetus of the field, starting with Shor's original factoring algorithm. A large enough fault-tolerant machine could break many classical cryptographic protocols. As a result, NIST has issued new standards moving encryption toward Post-Quantum Cryptography (PQC).

Hamiltonian simulation remains amongst the most explored and compelling applications for quantum computing. This problem of simulating complex quantum mechanical systems was the original inspiration for quantum computing. Though classical methods, including tensor networks, have found new success in simulating complex quantum systems, it is believed that QCs are the only way to scalably simulate large quantum systems.

**Architectural and algorithmic innovations are required to demonstrate applicability**

Unfortunately, resource estimates of both quantum factoring and Hamiltonian simulation suggest substantial architectural adjustments are required to achieve the first quantum speedups (Beverland et al., 2022). To address these overheads, increased focus must be placed on reducing resource requirements of these algorithms and co-designing algorithms with architectures. Current algorithmic innovations prioritize asymptotic advantages over resource efficiency of fault-tolerant hardware. We recommend algorithmic and architectural studies to reduce the constant factor resource requirements of these algorithms, rather than focusing on asymptotic speedups.

## 2.2 Exploring interdisciplinary applications could yield near-term results

Though keystone applications present compelling theoretical evidence for exponential advantage, the field will likely require near-term applications to sustain momentum. Developing near-term applications would justify the belief that QCs eventually may yield advantage.

We recommend an interdisciplinary approach to develop diverse applications. The field has primarily explored applications in physics (e.g. Hamiltonian simulation) and computer science (e.g. data analysis and processing). Identifying new applications for near-term quantum devices will require active engagement and dialogue with domain experts. Additionally, quantum-inspired algorithms which emerge from QIS research can be considered tangible deliverables from the field.

**Developing real world applications requires domain engagement**

Real-world applications require domain-specificity to appeal to practitioners and be adopted as feasible solutions. For example, problems in machine learning, optimization, and chemistry require deep integration with domain sciences to be effective. Quantum algorithms in these areas, while still requiring rigorous theoretical justification, have shown increasing algorithmic innovation and a steady stream of demonstrations on a variety of quantum hardware platforms. This work could be significantly accelerated by rigorous benchmarking (described in section 2.3) and deeper integration with domain experts to formulate realistic problems and constraints.

**Quantum-inspired classical algorithms are tangible deliverables from quantum research**

Furthermore, quantum-inspired classical algorithms may emerge from the pursuit of quantum algorithms. Recent work in quantum-inspired linear algebra (Chia et al., 2022) has shown how quantum-inspired algorithms can yield immediate benefits on classical devices while also facilitating a long-term transition to more advanced quantum algorithms. These classical algorithms should be seen as a deliverable which emerges as a result of concerted efforts in the field of quantum computing.

**Studying near-term applications can provide inspiration for long-term applications**

Encouraging the exploration of near-term applications can move us towards solving complex problems on existing and larger devices. For example, the search for near-term algorithms and QC prototyping can lead to novel quantum-inspired classical algorithms in addition to advancing the field of quantum computing. Thus, we recommend continuing to expedite near-term quantum computing applications and algorithms.

New algorithms and paradigms can emerge from the interaction between end users and quantum systems, for example: domain-specific approaches to compress classical data within QCs, adapting embedding techniques used in quantum chemistry to QCs, or modeling multivariate distributions with maximally entangled quantum states. Therefore, deeper integration between domain specialists and quantum algorithmicists may enable the discovery of new applications. Developing quantum intuition for scientists across disciplines could occur via visioning workshops with QC scientists and experts in other domains or hosting interdisciplinary conferences where end users are explicitly encouraged to collaborate.

## 2.3 Benchmarks facilitate application and algorithm development

Existing benchmark suites (e.g., SuperMarQ (Tomesh et al., 2022), QED-C (Lubinski et al., 2023)) have advanced the field by quantifying the capabilities of hardware (Li et al., 2022). Further work is necessary to incorporate new algorithms and capture the end-to-end performance of fault-tolerant algorithms (Lubinski et al., 2023). To understand the performance of quantum systems, and to demonstrate the potential benefits of QCs, the field requires compelling and standardized evidence to evaluate domain problems, algorithms, and hardware platforms. Rigorous benchmarking frameworks would enable principled comparison of platforms and solutions, as well as communicate the potential of these devices to the broader scientific and general communities.

We recommend exploring standardized benchmarking frameworks to identify a set of benchmarks which would enable us to evaluate quantum platforms, algorithms, and potential domain problems. For example, an end-to-end quantum machine learning benchmark would allow us to evaluate not only the general performance of a quantum device, but also the algorithm's noise resilience and data sensitivity. More work on widely accepted benchmarks with input from other communities (computer scientists, machine-learning communities) may also lead to increased collaboration and interest from other domain experts.

**Acute need to develop benchmarking methodology**

Further work is needed to holistically evaluate and compare quantum systems with different applications and algorithms. For example, current benchmarks do not allow us to consider the utility of quantum devices when combined with error mitigation techniques (Kim et al., 2023). To move toward more holistic and application-oriented benchmarks, we need to answer the following questions:

◗ What metrics should be used?

◗ What standards are necessary when evaluating QCs on these benchmarks?

◗ How should scaling studies be performed?

It is likely that no individual benchmark will holistically capture a device's performance, i.e., the above requirements likely cannot be satisfied simultaneously. Together, our benchmarks should holistically evaluate the entire stack, including how quantum algorithms compare to their classical counterparts when solving domain problems, studying hardware performance at running different error correction codes and hosting algorithms, and evaluating end-to-end performance (including the cost of running classical control hardware necessary for error correction).

**Evaluating subroutines**

Separately, there are algorithms that must be developed, which serve different purposes from demonstrating quantum advantage, such as those that can test for the use of quantum operations, and algorithmic kernels that help to explore key machine attributes such as connectivity, and the role of mid-circuit measurements. Such kernels will also serve to inspire next-generation machine designs.

## 3. Fault Tolerance and Error Mitigation

For QCs to be useful they need to be reliable. Quantum systems have improved in reliability over the last five years but to build QCs with the capability of millions or billions of operations more work is needed. Error mitigation is a broad class of methods for estimating the noise free output of a QC from a diverse set of noisy computations. QEC encodes quantum information so that errors can be detected and fixed. Fault-tolerant quantum computation is built off QEC and is currently the best known path for enabling billions or more operations, but the community is at the start of this path. In this section, we describe opportunities and challenges of achieving fault-tolerant QCs.

### 3.1 Fault-tolerant methods are required for large-scale quantum computers

Quantum fabrics have advanced greatly over the last five years exhibiting both increased qubit numbers and improved gate quality. Although physical quantum gates are continuously improving, it is challenging to imagine the physical gates failing at a rate below one-part-per-million. Many large scale quantum algorithms such as Shor's factoring algorithm and high-precision quantum simulations are expected to require billions or trillions of gates to be competitive with standard computers (Gidney et al., 2021). The theory of fault-tolerant quantum computation describes a path to reach these low errors, but some hurdles stand in the way of applying this theory to physical quantum systems.

The challenge of fault-tolerance is the overhead requirement in terms of qubits and gates. The surface code is a promising platform for fault-tolerant QEC and can be built on systems with a planar connectivity graph and boasts a high error threshold (Raussendorf at al., 2007). The surface code does not densely encode information and the code rate vanishes in the asymptotic limit. In the last five years, there has been great progress on finite-rate QEC codes with low-weight parity checks (qLDPC codes) (Breuckmann et al., 2021). These codes can now achieve linear rate and distance. The downside is these codes require very non-planar connectivity and questions about parallel computation remain.

It is imperative that the quantum computing community continues to develop larger, scalable, and affordable systems that can absorb the cost of quantum error correction. The computing community can also provide valuable insight for reducing the overhead and programming these future devices.

### 3.2 Co-design of QEC, quantum hardware, and classical hardware is necessary

**Rethink device topology to enable new codes:** This is a call to arms for developing quantum hardware that supports greater connectivity. Current 2D connectivity on a single plane is sufficient for implementing surface codes (Google Quantum AI, 2023) but greatly limits the potential to develop more efficient code-to-hardware mappings. However, more efficient codes, such as general low-density parity check (LDPC) codes, require non-local connections that are difficult to map on current architectures. These could be achieved practically if the connections could be made, for example, over multiple planes. This is much like what is done in conventional chips, where different signals are routed over 8-10 layers above the silicon layer, enabling long-range (non-local) connections. Such an architecture could be tailored to specific codes, such as LDPC codes or other more efficient codes. As we move to codes with larger distance, the area overhead of surface codes may become prohibitive (due to d-square requirement on the number of physical qubits), and enabling the hardware to support LDPC codes may be a more practical way to enable stronger fault tolerance efficiently.

**Decoding, real-time, scalable, and morphable:** At the device level, the constraints on error correction vary depending on the qubit technology. For example, trapped ion systems can tolerate longer decoding latencies compared to superconducting systems. Similarly, the way surface codes are mapped and decoded on a grid lattice (like the Google devices) differs from a heavy hexagonal lattice (such as the more recent IBM architectures). On the other hand, the code redundancy requirements may vary depending on the application. Moreover, the landscape of error correction is evolving rapidly with the emergence of newer fault modes. For instance, leakage errors or errors from cosmic rays striking the qubits severely impact the performance of error correction. Overall, each combination of applications of QEC code and quantum hardware present their own unique set of constraints, and the control and decoding architecture must adapt to tackle them. Thus, decoders will not only need to be real-time, but also morph between different code-types and distances for even the same machine (Battistel et al., 2023).

**Improving QEC codes through tailored noise:** Recent work has shown the advantage of designing qubits to achieve a certain type of noise for QEC. Cat-codes can be constructed that are bias-noise preserving, leading to a large discrepancy between phase flip noise (Z) and bit flip noise (X). Schemes have been developed for neutral atoms, ions, and superconductors that make erasure errors the dominant error. Although erasure errors completely remove the local information of the state, they provide additional information by pointing out which qubit has gone wrong.

### 3.3 Resource estimation tools needed to evaluate future devices

Co-designing QEC with the quantum device, classical hardware, and software requires evaluating the impact of design decisions at scale. Yet without large scale devices today, enabling parallel progress across these domains requires modeling of the full quantum computing stack. To this end we need comprehensive and flexible quantum resource estimation tools, which provide a common means of comparison. These tools will need to be modular, allowing researchers to test different layers of the FTQC stack, and explicit, ensuring impactful assumptions are clearly stated. By creating these tools, researchers can both explore solutions to existing problems and identify new problems that appear when moving from theory into practice.

### 3.4 Error mitigation and error correction techniques can complement each other

Error mitigation may provide a path towards commercially viable quantum simulation applications. In a certain sense, this path could conceivably lead to QEC by, for example, simulating a toric code. Along this path we expect to make progress towards noise characterization in NISQ systems.

In addition, techniques from fault-tolerant circuit design can find application in detecting and/or correcting certain errors before full fault-tolerance is achieved, and vice-versa, error mitigation on parts of the QEC protocol can help relieve some of the resource requirements of these protocols.

### 3.5 Fault-tolerant systems and computer architecture

Fault-tolerant systems provide different architectural and systems challenges compared to NISQ era systems. In the NISQ era, there is a wider variability in qubit quality and an extended set of machine operations including pulse-level control. For fault-tolerant QCs, the code provides protection but also yields a simplified set of control instructions. These instructions are split between easy to implement transversal gates and more challenging gates that need to be injected into the code to achieve universality. To increase the quality of injected states magic-state distillation is often required. In the NISQ era, two-qubit gates are often the bottleneck. For the fault-tolerant era, arbitrary single-qubit gates become challenging. Optimization work in the past has focused on a limited set of codes assuming a homogeneous computational fabric. Future work will examine modules, concatenations of bosonic and stabilizer codes, and heterogeneous qubit designs.

# 4. Hybrid Quantum-Classical Systems: Architectures, Resources, Management, and Security

Since quantum algorithms and hardware have thus far served as special-purpose accelerators, the solution of most practical problems will require substantial classical computation in cooperation with a quantum kernel. Indeed, we've already seen this relationship define the performance capabilities of today's devices. Classical pre-processing for circuits and post-processing for outputs to boost program fidelity are critical for near-term quantum success. Even in the regime of fault tolerance, these techniques will be necessary to keep error rates low enough to apply quantum error correcting codes. Beyond the optimization of program performance, however, variational algorithms constitute a setting where classical computation is applied in alternation with quantum kernels, and error correction decoding requires a constant, low-latency quantum-classical loop to keep up with errors in real time. All of these settings require varied classical systems in conjunction with the quantum device, each having different metrics of performance.

## 4.1 Classical optimization

Classical optimization is vital to improving the quality of quantum applications, requiring state-of-the-art techniques. Matching algorithms written in a growing body of quantum programming languages to the highly varied set of primitive hardware operations requires constantly evolving circuit synthesis techniques and could benefit from intermediate representations. Optimizations need to then map qubits and route operations of synthesized circuits while respecting a hardware's architectural constraints, such as a limited connectivity between physical qubits. Additionally, the fidelity of the executed quantum circuit can further be improved by intelligent post-processing of measurement results.

Although classical techniques for circuit synthesis, mapping, and routing have been explored for near term quantum applications, they are equally important in a fault tolerant setting. To preserve fault tolerance, logical operations are constrained to those that are protected by the quantum error correcting code, creating a distinct circuit synthesis problem. Performing operations between encoded qubits is also distinct; for example, operations between surface code qubits using lattice surgery requires allocating a path of ancilla connecting the qubits. Answers to these optimization problems have only recently been explored, and will be key as hardware progresses towards QEC. However, optimizations at the physical layer are still important for QEC. Since quantum error correcting codes require physical error rates to be below some threshold, well designed classical techniques can accelerate reaching requisite error rates. For example, hardware-tailored optimizations have been employed in early experimental demonstrations of QEC (Google Quantum AI, 2023).

It's important to note that all of these optimization problems are computationally hard. As we move to larger quantum systems, the scalability of deep optimization techniques becomes challenging: circuits for applications with known quantum advantage are estimated to require 10s of millions of qubits and billions of gates (Gidney et al., 2021). As such, it is imperative to design optimization techniques that balance a full-stack approach versus an abstraction-based approach. Achieving this balance requires inputs from both the application requirements as well as technology constraints.

## 4.2 Variational algorithms

Many promising algorithms, particularly for the near-term, are variational: using a classical optimizer with a quantum kernel. As hybrid quantum-classical systems, these algorithms have unique challenges. For example, noise in the quantum device can cause barren plateaus in the cost landscape, requiring exponentially many shots which can remove any potential advantage. Errors in the quantum device can also drift over time, making optimization difficult. These challenges are exacerbated by the fact that most QCs today are available as cloud services, where large numbers of shots can be costly and limited control over hardware scheduling can increase drift during optimization. In light of these challenges, there are multiple avenues current and future research can explore. Hardware-tailored error mitigation techniques can alleviate the prevalence of barren plateaus, and combined with quantum-aware optimizers, reduce necessary shot costs. Novel tools and programming languages can also improve the development and execution speed of variational algorithms, closing the gap to a potential demonstration of quantum advantage.

Taking a step back, a key perspective on these first two hybrid systems is that they can be a contest of two exponentials – in which we can invest up to an exponential amount of classical processing to enable quantum computations that hopefully have exponential advantages over classical alternatives. This investment is worthwhile as long as the overall hybrid quantum cost is lower than the fully classical cost.

## 4.3 Simulation, verification, and debugging

Hybrid quantum-classical systems also encompass the areas of simulation, verification and debugging. Simulation is a vital tool, both to verify the capabilities of quantum hardware and to provide noise-free computational support to the quantum device. For example, circuits made from only Clifford gates are known to be efficiently simulatable. As a result, viewing Clifford circuits as a quantum tool yields no advantage, however, usefulness can be found when viewing Clifford circuits as a classical tool. Specifically, classical simulation of Clifford circuits can give better initial states for variational algorithms, bootstrapping algorithmic performance (Ravi et al, 2022).

Classical simulation can also play a large role in the debugging of quantum applications, which is particularly challenging due to the black box nature of quantum computing. Adaptation of compressed sensing techniques have shown efficient ways to debug variational algorithms (Liu et al., 2023), and future research will need to continue this theme of quantum-classical debugging.

Notable experimental quantum demonstrations also give rise to improved classical simulations. Existing claims of quantum supremacy have been matched by novel classical techniques, and as quantum hardware matures, such innovations in classical simulations will further refine our understanding of the capabilities of classical and quantum computation.

## 4.4 Decoding for QEC

With the increasing shift of focus towards fault tolerant quantum computation, an important problem is decoding quantum error correcting codes via a classical computer. This is notably different from other hybrid systems in that the classical decoder must operate in real-time with the quantum device which is inherently limited by qubit coherence. The result is a setting with high-bandwidth requirements at the quantum-classical interface. For example, fast quantum devices such as superconducting, photonic, and topological devices can produce decoding data at rates of Gbps for each encoded qubit. A classical system unable to keep up with this data runs into a backlog, causing an exponential slowdown (Terhal 2015). Furthermore, decoding algorithms are not constant in runtime, but instead scale with the code size. At sizes necessary for advantageous algorithms, the classical time spent on decoding can be prohibitively large, creating the dreaded backlog. Recent work has looked at this problem for surface codes at both the algorithmic and classical systems level, but further research needs to be done to address these issues for broader codes such as quantum LDPC codes. Many quantum devices also operate in cryogenic refrigerators, creating further difficulties for real-time processing and decoding. Nonetheless, these classical systems will need to be developed as fault tolerant quantum computing comes closer to reality. One path towards innovation is using academic testbeds for experimentation on algorithms and control systems, including exploration of hardware-software trade offs.

## 4.5 System security

An emerging area is identifying securities and vulnerabilities that affect quantum systems. Despite the small scale of current devices, it's important to look at vulnerabilities with an eye towards addressing them from the beginning, before getting tied into a model. Because classical programs act as the gateway to quantum systems, a large section of the security risks can be mitigated through the use of security protocols from classical computing systems. However, there is a need to focus on security risks that are specific to quantum systems. An example of such vulnerabilities is cross-talk errors when sharing a quantum system to run multiple workloads, which can be tackled through the use of simple techniques like physical distancing between partitioned qubits. It is also vital to be cautious about critical control software and compiler code, as malicious low-level access on quantum systems can be damaging.

## 5. Tools and Programming Languages

Quantum programming today requires a deep knowledge of unitary mathematics and its associated linear algebra. Even with this knowledge, well-known algorithms are non-intuitive to newcomers, and, new algorithms are difficult to reason about even for quantum experts. To welcome newcomers to the field, to facilitate research, and to permit scaling up to programs with quantum advantage, efficient high-level quantum programming abstractions are needed. To realize such abstractions, software engineering infrastructure is needed for compilation, verification, and simulation, both for near-term and long-term hardware.

### 5.1 The need for higher-level languages

The knowledge needed to develop new algorithms for QCs presents a high barrier to entrance, and this represents a significant challenge for quantum computing adoption. Anecdotal reports include quantum computing classes at universities spending so much time on the underlying physics that little time is left to cover quantum computing beyond well known algorithms such as Grover's (1996) and Shor's (1999). Although there exists several quantum programming languages today (e.g., Q# (Svore et al., 2018), Scaffold (Abhari et al, 2012) (Litteken et al., 2020), Qiskit (Qiskit Contributors, 2023), OpenQASM (Cross et al., 2017) (Cross et al., 2022), etc.), virtually all existing quantum languages are in essence different ways of expressing the interconnections between quantum gates. It is difficult to imagine a higher-level abstraction.

We can reach back to the early days of digital logic and computer technology to find similarities between then and now. Relay-based switching circuits were engineered with their own set of principles until Claude Shannon, among others, noticed the isomorphic relationship to Boolean algebra (Shannon, 1938). After this revelation, a large body of work became applicable to the new field. Similarly, the early days of Integrated Circuits (ICs) required physicists and electrical engineers to innovate. Mead and Conway (1980) established the lambda grid and a series of design rules that enabled a much larger community to design ICs, leading to large scale and very large scale circuits including the early RISC microprocessors.

The primary research question is whether there exists one or more new abstractions to express quantum computation that will be comprehensible to a much broader audience than today and, if so, how will these new languages be discovered. At the time of this writing, there are but a handful of proposals of varying degrees of generality. Many more proposals must be brought to the table. It is also important to note that there need to be at least two classes of such languages: those designed for the era of fault-tolerant systems with plentiful qubits, and those for NISQ systems (see section 5.3 below). Both need to be developed, and there should not be an emphasis on one at the expense of the other. The barriers to success for both of these classes of languages include the lack of academic venues for debate and the resources needed to stand up these efforts.

Academic venues today are concentrated on classical programming languages. The same problem with teaching quantum computing without first teaching quantum mechanics exists in these communities. Getting drafts reviewed for publication becomes difficult if not impossible. Part of the reason is the small size of the QC community. But at the same time, without sufficient peer-reviewed validation of ideas, the community will not be able to grow.

Coupled to the dearth of academic venues is the problem of fueling new research in the area of QC languages. Funding agencies that peer-review proposals suffer from the same problems with evaluating this work. Without a conscious effort to overcome these barriers, the new research will not proceed.

Thus what is needed in this space of QC is very similar to other areas of QC research: a spark to start the community. This will require some risk. However, it will not get off the ground without significant investment and encouragement.

### 5.2 Quantum multi-level intermediate representations

There is a need for multi-level intermediate representations (IRs) that can carry high-level semantics through multiple optimization stages, encode all known algorithm constraints, and be compatible with further low-level optimizations. We highlight a few design philosophies for exploiting multi-level quantum IRs. First, expressiveness is crucial and an IR should be capable of efficiently encoding large computation blocks for quantum computing, e.g., an arbitrary n-qubit gate, without the need to write down verbose matrices explicitly. Second, the IR should be universal, with the power to support any quantum program, and allow for flexibility in the types of computations that can be represented. Thirdly, to facilitate seamless compilation, we need an easy lowering process from a high-level IR to a lower-level IR. For instance, this can involve mapping high-level operations to a hardware-native gate set like 1-qubit gates and the CNOT gate, or to a fault-tolerant gate set like Clifford+T.

Multi-level IRs, designed with the recommendations above, offer significant benefits. They enable large-scale optimizations that are often hard to achieve at pure lower IR levels. By operating at multiple levels of abstraction, the IRs open up possibilities for extensive optimizations, leading to improved overall performance and efficiency in quantum compilation.

The IRs described above are not the only need in this space. Drawing inspiration from LLVM (Low Level Virtual Machine) (Lattner et al., 2004), a compiler framework that has united classical compiler research, industrial development and productization, the QC community needs a large, funded effort to create a framework for realizing the IRs. This framework cannot be created in a vacuum. Rather, it must be created as a collaboration, grounded in academia, but closely tied to and responsive to the needs of industry. One aspect that has made LLVM a success is this crosscollaboration between academic and industrial development. The IR framework must equally support research and have forks or branches that are hardened for industrial applications. However, the competition between QC technologies existent in the industry today is critical. Without careful shepherding, the IR framework could accidentally pick winners and losers. We recommend a steering committee composed of all shareholders that acts as a standards body to guarantee against this hazard.

## 5.3 The challenges of NISQ machine programming

Achieving quantum advantage on noisy near-term hardware requires taking full advantage of device physics. Consequently, even for promising NISQ applications such as quantum chemistry, exploiting hardware characteristics still requires domain experts (e.g., chemists) to have a deep knowledge of hardware strength and limitations. Programming languages and tools will abstract these hardware details away while still achieving comparable performance.

One way this could be achieved is through identifying core abstraction building blocks. These building blocks can be specialized to hardware and application specifics through collaboration between quantum computing engineers and subject-matter experts, and be utilized by domain experts without having to break the abstraction barrier. Identifying these building blocks and characterizing the opportunities for specialization is an important step forward in enabling consistent use by non-quantum computing experts.

On the other hand, in the near term, many users of quantum machines will in fact be quantum computing experts and quantum physicists looking to explore and exploit the capabilities of the machines. QCs are physically interesting devices, and we need to provide ways for physicists to interact with these devices and learn about their behavior: their decoherence times, levels of crosstalk, SPAM (state preparation and measurement) error, and error rates across different gates and other operations. These metrics can inform us of the physical limits of the underlying technologies.

Likewise, quantum computing experts need to be able to directly interact with the hardware in order to stretch it to their applications. A variety of approaches have been taken here, from using the extra energy states in transmons to represent qutrits or higher-dimensional qudits to synthesize entangling gates for specific qubits on specific hardware (Lin et al., 2022). This access to the hardware will be invaluable for the foreseeable future, meaning that we need languages that allow programmers arbitrary access to the quantum devices, including pulse-level control and readouts. This should not mean full device-level control, however, which could endanger the devices themselves. Such control would also be a burden to the programmers, who should not need to learn a new language for every device they interact with. Instead, we need domain specific languages that facilitate low-level interaction with a variety of quantum devices. Such abstract analog instruction sets (Alexander et al., 2020) (Peng et al., 2023) will likely prove a powerful aid to quantum computing and quantum engineering experts for the foreseeable future.

## 5.4 Verification, testing and validation across the software stack

Quantum compilers and tools have the potential to introduce bugs into quantum programs, which are among the hardest to understand and fix in practice. Below, we outline some tools and techniques that can help avoid introducing those bugs in the first place.

One common technique is *formal verification*, which can be used to mathematically show that a program matches a given specification using tools ranging from theorem provers to SMT solvers.

Other lightweight techniques for increasing reliability of quantum tools include unit tests, runtime assertions and random testing. The applicability of these tests will vary based on the application. For example, quantum circuit optimizers take in a certain pattern and transform it into a more succinct, equivalent pattern. In this setting, tools can test or verify the property that output patterns are semantically equivalent and shorter to the input, and analyze coverage of those tests by considering the input space of possible circuits. Similar techniques can be applied to (logical) quantum circuit programs, where we have concrete inputs and expected outputs, leveraging the existing literature.

Currently, approaches applying formal verification to quantum computing have focused on circuit-level tools, including compilers and programs. While these efforts should continue and plenty of work is left to be done at this level of abstraction, more work is needed at lower abstraction levels. Circuit-level verification models idealize the quantum computing stack, assuming logical qubits and perfect quantum gates. But verification is also crucial for the more concrete layers of the stack, and requires specialized knowledge about target architectures, including accurate error models.

One prime verification target is the control software governing quantum pulses. Such control systems tend to be designed from scratch for each quantum device, and are a frequent source of bugs: in particular, concurrency bugs. As an example, one could aim to verify that distinct pulses do not overlap, which guarantees that certain kinds of crosstalk don't occur. Verifying some properties may also be purely classical, and can take advantage of classical verification techniques. Other properties may need to develop new techniques based on quantum semantics.

Things become more complicated when we think about real, error-prone QCs, and even more so when we think about the continuous analog pulses that govern their behavior. Here it becomes hard to specify the behavior of the program and to test in such a way that potential sources of error will be covered. Specification is still possible, but it has to be probabilistic. In the case of reasoning about the pulse level, things become even fuzzier, with a continuous range of possible outcomes.

## 5.5 Incorporating realistic hardware constraints into tools and simulations

We need to build simulators, resource estimators, and other analysis tools that incorporate realistic hardware constraints at a variety of abstraction levels, for example noise models. Such tools are crucial for both architects and experimentalists, who use simulation to guide the design of hardware and control electronics. At the other end of the stack, domain experts use resource estimation tools to determine the feasibility of quantum use-cases. Both sets of tools would benefit from incorporating realistic noise models and architectures.

We envision a few key principles for guiding the design of long-lasting tools for benefiting the field of quantum computing. First, it is crucial to incorporate hierarchical multi-granularity in the design of these simulation tools. Hierarchical multi-granularity refers to the ability to model and simulate different levels of abstraction in a quantum computing system. For example, at a lower level of granularity, we may require Electronic Design Automation (EDA) tools to guide qubit engineering. These tools can simulate the physical characteristics and behaviors of individual qubits, allowing for optimization and refinement of their performance. At a higher level of granularity, the simulation tools should be able to model the mapping and execution of the entire quantum circuits to hardware, taking into account the coupling graph and noisy model of the hardware.

Secondly, a key requirement for these simulation tools is modularity, allowing for the reuse of components across different quantum hardware and technologies. The quantum computing landscape is characterized by diverse hardware types, including different encoding (e.g., qubits, qudits, continuous variables), quantum technologies (e.g., superconducting, ion trap, neutral atom), and programming paradigms (e.g., gate-based, measurement-based, analog computing). This diversity necessitates the development of modular and flexible simulation tools that can accommodate and optimize across this wide range of possibilities, enabling researchers to explore and compare performance across different platforms and paradigms. Such tools are vital for advancing quantum computing towards practical applications.

## Conclusion

Quantum computing is at a historic and pivotal time, with substantial engineering progress in the past 5 years and a transition to fault-tolerant systems in the next 5 years. Taking stock of what we have learned from NISQ systems, this report examined 5 key areas in which computer scientists have an important role in exploring. These areas are:

### 1) Technologies and Architectures with a View Towards Scaling

Needs include lower computational error and decreased cost per qubit. Additionally, new models that help understand the dynamics of actual systems and their constraints are necessary. Achieving this will require combined efforts of academia and industry. New funding is needed to catalyze these relationships.

### 2) Applications and Algorithms

Quantum computing advances are limited by the difficulty in creating new applications and algorithms to demonstrate a practical quantum advantage. At the same time, key existing applications need investment to move from theoretical to practical implementation. The resource requirements for keystone applications need to be reduced. Also, there is a need for both domain integration for near-term applications, as well as hardware benchmarking to encourage more algorithm development.

### 3) Fault Tolerance and Error Mitigation

Noise is the current limitation to quantum computing. In the near-term, there needs to be more research into error mitigation and QEC. Achieving large scale quantum computing will require error correction and fault tolerance. Co-design by quantum architects and coding experts is needed to discover new QEC codes and to bridge between current systems and future large-scale systems.

### 4) Hybrid Quantum-Classical Systems: Architectures, Resource Management, and Security

Many practical problems require a hybrid solution between quantum and classical systems. Research is needed into the design of such hybrid systems, the applications to run on them, and the classical computation for quantum circuit optimization, simulation, and verification. In addition, how to secure quantum systems against potential vulnerabilities is an emerging issue that remains to be addressed.

### 5) Tools and Programming Languages

The level of expertise required to program a quantum system today is very high and requires advanced, post-graduate education. This barrier to entry makes it difficult to expand the field beyond the current community. Efficient, high-level quantum programming languages and program abstractions are desperately needed. These necessarily require infrastructure for compilation, verification, and simulation, for both near-term and future quantum hardware.

It is our hope that the directions discussed here will energize our community and advance a systems approach to quantum computing that will help this important computing paradigm move forward towards practical applications.

5 YEAR UPDATE TO THE NEXT STEPS IN QUANTUM COMPUTING REPORT## References

1. Abhari, A. J., Faruque, A., Dousti, M. J., Svec, L., Catu, O., Chakrabati, A., Chiang, C.-F., Vanderwilt, S., Black, J., Brun, T., Pedram, M., Brown, K., Suchara, M., Martonosi, M., & Chong, F. (2012, July 24). *Scaffold: Quantum Programming Language*. Princeton University. https://www.cs.princeton.edu/research/techreps/TR-934-12

2. Acharya, R., Aleiner, I., Allen, R., Andersen, T. I., Ansmann, M., Arute, F., Arya, K., Asfaw, A., Atalaya, J., Babbush, R., Bacon, D., Bardin, J. C., Basso, J., Bengtsson, A., Boixo, S., Bortoli, G., Bourassa, A., Bovaird, J., Brill, L., … Zhu, N. (2022, July 20). *Suppressing quantum errors by scaling a surface code logical qubit*. arXiv.org. https://arxiv.org/abs/2207.06431

3. Aleksandrowicz, G., Alexander, T., Barkoutsos, P., Bello, L., Ben-Haim, Y., Bucher, D., Cabrera-Hernández, F. J., Carballo-Franquis, J., Chen, A., Chen, C.-F., Chow, J. M., Córcoles-Gonzales, A. D., Cross, A. J., Cross, A., Cruz-Benito, J., Culver, C., González, S. D. L. P., Torre, E. D. L., Ding, D., … Zoufal, C. (2019, January 23). *Qiskit: An open-source framework for quantum computing*. Zenodo. https://zenodo.org/record/2562111

4. Alexander, Thomas, et al. "Qiskit pulse: programming quantum computers through the cloud with pulses." Quantum Science and Technology 5.4 (2020): 044006.

5. Battistel, F., Chamberland, C., Johar, K., Overwater, R. W. J., Sebastiano, F., Skoric, L., Ueno, Y., & Usman, M. (2023, May 22). *Real-time decoding for fault-tolerant Quantum Computing: Progress, challenges and outlook*. arXiv.org. https://arxiv.org/abs/2303.00054

6. Beverland, M. E., Murali, P., Troyer, M., Svore, K. M., Hoefler, T., Kliuchnikov, V., Low, G. H., Soeken, M., Sundaram, A., & Vaschillo, A. (2022, November 14). *Assessing requirements to scale to practical quantum advantage*. arXiv.org. https://arxiv.org/abs/2211.07629

7. Breuckmann, N. P., & Eberhardt, J. N. (2021). Quantum low-density parity-check codes. *PRX Quantum*, *2*(4). https://doi.org/10.1103/prxquantum.2.040101

8. Chia, N.-H., Gilyén, A. P., Li, T., Lin, H.-H., Tang, E., & Wang, C. (2022, October 1). *Sampling-based sublinear low-rank matrix arithmetic framework for Dequantizing Quantum Machine Learning*. Journal of the ACM. https://dl.acm.org/doi/abs/10.1145/3549524

9. Cross, A. W., Bishop, L. S., Smolin, J. A., & Gambetta, J. M. (2017, July 13). *Open quantum assembly language*. arXiv. http://arxiv.org/abs/1707.03429

10. Cross, A., Javadi-Abhari, A., Alexander, T., De Beaudrap, N., Bishop, L. S., Heidel, S., Ryan, C. A., Sivarajah, P., Smolin, J., Gambetta, J. M., & Johnson, B. R. (2022). OpenQASM 3: A broader and deeper quantum assembly language. *ACM Transactions on Quantum Computing*, *3*(3), 1–50. https://doi.org/10.1145/3505636

11. Gidney, C., & Ekerå, M. (2021). How to factor 2048 bit RSA integers in 8 hours using 20 million noisy qubits. *Quantum*, *5*, 433. https://doi.org/10.22331/q-2021-04-15-433

12. Google Quantum AI. (2023, February 22). *Suppressing quantum errors by scaling a surface code logical qubit*. Nature News. https://www.nature.com/articles/s41586-022-05434-1

13. Grover, L. K. (1996). A Fast Quantum Mechanical Algorithm for Database Search. *Proceedings of the Twenty-Eighth Annual ACM Symposium on Theory of Computing - STOC '96*, 212–219. https://doi.org/10.1145/237814.237866

14. Kim, Y., Eddins, A., Anand, S., Wei, K. X., van den Berg, E., Rosenblatt, S., Nayfeh, H., Wu, Y., Zaletel, M., Temme, K., & Kandala, A. (2023, June 14). *Evidence for the utility of quantum computing before Fault Tolerance*. Nature News. https://www.nature.com/articles/s41586-023-06096-3

15. Lattner, C., & Adve, V. (2004). LLVM: A compilation framework for lifelong program analysis & transformation. *International Symposium on Code Generation and Optimization, 2004. CGO 2004.*, 75–86. https://doi.org/10.1109/cgo.2004.1281665
16

## Workshop participants

| First Name | Last Name | Affiliation |
|---|---|---|
| Austin | Adams | Georgia Tech |
| Aydin | Ayanzadeh | University of Maryland, Baltimore County |
| Ramin | Ayanzadeh | Georgia Tech |
| Kenneth | Brown | Duke University |
| Gregory | Byrd | NC State University |
| Fred | Chong | University of Chicago |
| Tom | Conte | Georgia Tech |
| Andrew | Cross | IBM Research |
| Dilma | Da Silva | NSF/CISE/CCF |
| Aniket | Dalvi | Duke University |
| Poulami | Das | Georgia Tech |
| Yi | Ding | Massachusetts Institute of Technology |
| Yongshan | Ding | Yale University |
| Yufei | Ding | UCSB |
| Tom | Draper | Center for Communications Research at La Jolla |
| Catherine | Gill | Computing Community Consortium |
| Peter | Harsha | Computing Research Association |
| Maddy | Hunter | Computing Community Consortium |
| Ali | Javadi-Abhari | IBM |
| Sonika | Johri | Coherent Computing |
| Christopher | Kang | University of Chicago |
| Katie | Klymko | Lawrence Berkeley National Laboratory |
| Olivia | Lanes | IBM Quantum |
| Nathalie | de Leon | Princeton University |
| Margaret | Martonosi | Princeton University |
| Anne | Matsuura | Intel Labs |
| Jennifer | Paykin | Intel |
| Raghavendra Pradyumna | Pothukuchi | Yale University |
| Moin | Qureshi | Georgia Tech |
| Robert | Rand | University of Chicago |
| Gokul Subramanian | Ravi | University of Chicago |
| Eleanor | Rieffel | NASA Ames Research Center |
| Taniya | Ross-Dunmore | Computing Research Association |
| Amr | Sabry | Indiana University |
| Ann | Schwartz | Computing Community Consortium |
| Kaitlin | Smith | Northwestern University and Infleqtion |
| Jakub | Szefer | Yale University |
| Swamit | Tannu | University of Wisconsin Madison |
| Jake | Taylor | Riverlane |
| Devesh | Tiwari | Northeastern University |
| Josh | Viszlai | University of Chicago |
| Robert | Wille | Technical University of Munich |
| Xiaodi | Wu | University of Maryland, College Park |
| Jon | Yard | University of Waterloo/IQC/PI |
| William | Zeng | Unitary Fund |

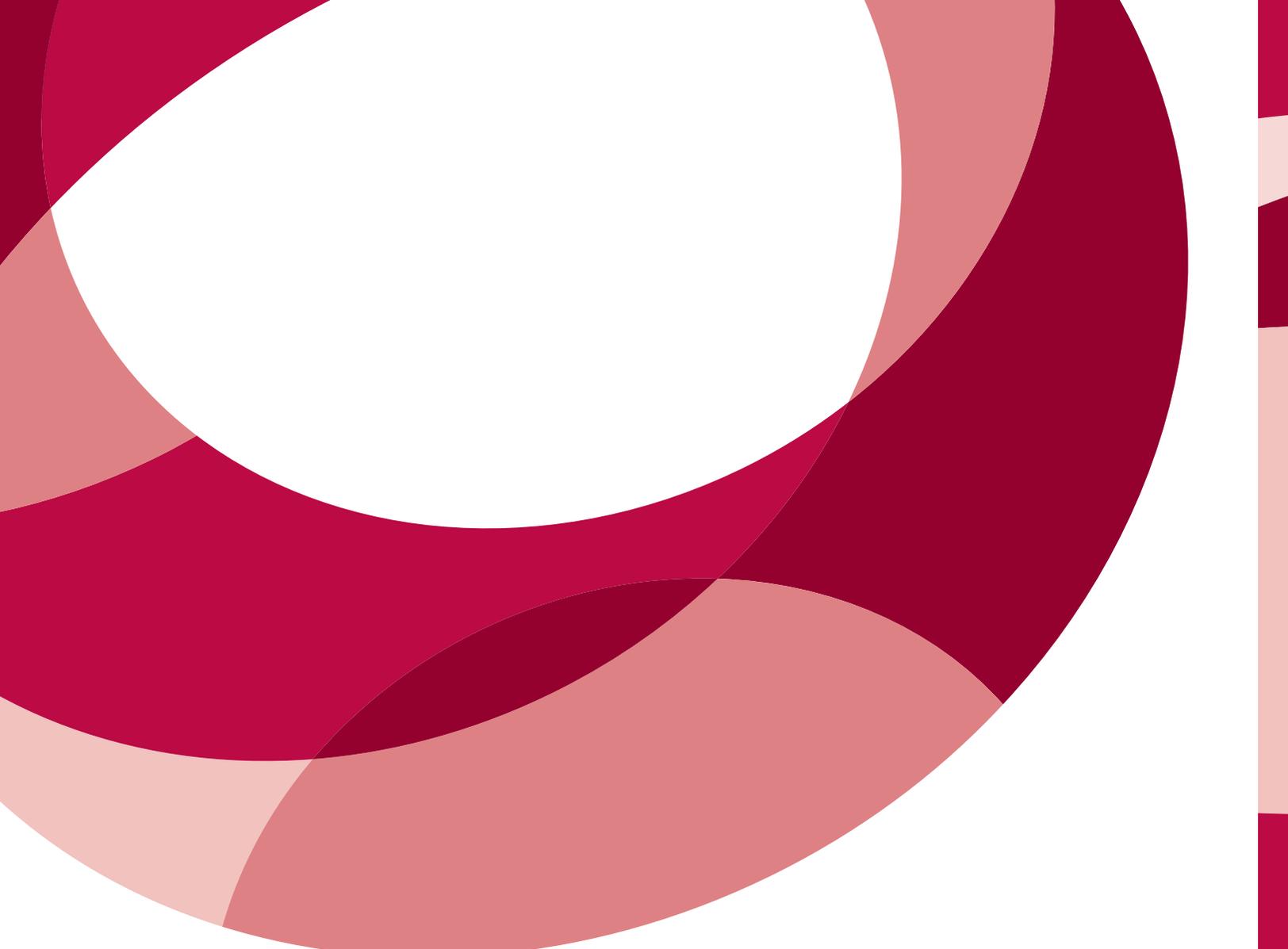

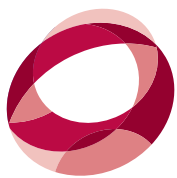

**CCC**
Computing Community Consortium
Catalyst

1828 L Street, NW, Suite 800
Washington, DC 20036
P: 202 234 2111 F: 202 667 1066
www.cra.org cccinfo@cra.org